\documentclass{PoS}

\title{Abelian dominance in local unitary gauges and without gauge-fixing in pure SU(2) QCD}

\ShortTitle{Abelian dominance in local unitary gauges and without gauge-fixing in pure SU(2) QCD}

\author{\speaker{Toru Sekido}, Katsuya Ishiguro, Yoshihiro Mori and Tsuneo Suzuki\\
        Institute for Theoretical Physics, Kanazawa University, Kanazawa
        920-1192, Japan\\ and RIKEN, Radiation Laboratory, Wako 351-0158, Japan\\
        E-mail: \email{toryu@hep.s.kanazawa-u.ac.jp}}

\abstract{
We perform lattice Monte-Carlo simulations 
of pure $SU(2)$ QCD using the multi-level method. We find Abelian dominance  in local unitary gauges such as those diagonalizing a plaquette. A static potential described by Abelian link fields alone 
gives us the same string tension as that of a non-Abelian potential.
Abelian dominance of the string tension and Abelian flux tube profiles are observed also without gauge-fixing, i.e., without any Abelian projection. On the basis of  these results,
we  propose a simple gauge-independent  Abelian confinement scenario without any Abelian projection.
All color components of the non-Abelian field strength  
become Abelian dominant  in the infrared region. The Abelian dual Meissner effect works in any color direction. Abelian neutral states  in any color directions which are just non-Abelian color-singlet can exist 
as a physical state. In this way, the non-Abelian color confinement could be understood in the framework of the Abelian dual Meissner effect. }

\FullConference{XXIVth International Symposium on Lattice Field Theory\\
                July 23-28, 2006\\
                Tucson, Arizona, USA}

\begin{document}

\section{Abelian dominance in Abelian projection.}

The Abelian dual Meissner effect is believed to be the color confinement mechanism in QCD
\cite{Mandelstam:1974pi}\cite{'tHooft:1974qc}.
In this model Abelian gauge fields and Abelian monopoles play 
very important roles. How to extract Abelian monopoles from 
non-Abelian QCD is a problem. 
In 1981 'tHooft suggested an Abelian projection method 
in which the Abelian degrees of freedom can be extracted \cite{tHooft:1981ht}.
A partial gauge-fixing is performed in the Abelian projection.
Only $U^{N_c-1}(1)$ subgroup remains unbroken after the Abelian projection.
On lattice, the Abelian gauge fields $\theta_{n,\mu}$ are defined by the following way:
(Here we consider the SU(2) case for simplicity.)
\begin{eqnarray}
U_{n,\mu}^{G}=C_{n,\mu}u_{n,\mu}, \\
u_{n,\mu}=\mbox{diag}(e^{\theta_{n,\mu}},e^{-\theta_{n,\mu}}),
\end{eqnarray}
where $U_{n,\mu}^{G}$ is a non-Abelian link fields after the partial gauge-fixing using
a gauge transformation matrix G.  
$C_{n,\mu}$ is an off-diagonal field.
The maximally Abelian(MA) gauge
,where link variables are abelianized as much as possible,
is the most famous gauge-fixing condition
in the Abelian projection.
Numerically the string tension calculated from 
Abelian parts reproduces well the original one 
in MA gauge\cite{Suzuki:1989gp}\cite{Bali:1998de}.
It is called as Abelian dominance.
There are many results which suggest the Abelian scenario 
in MA gauge(e.g. flux tube profiles\cite{Koma:2003gq}).
In MA gauge,  the Abelian scenario based on the Abelian dual Meissner effect works good.
However there is a serious problem in the Abelian scenario.
Namely good results suggesting the success of the  Abelian scenario are obtained 
with only a few non-local gauges which are similar to MA gauge.
Particularly the Abelian scenario has not been observed in 
local unitary gauges such as F12 gauge
\cite{Suzuki:1989gp}\cite{Bernstein:1996vr}\cite{Ito:2002zv}.
We call it 'gauge dependence problem of the Abelian scenario'.
In this short note, we investigate this problem
with the method of  high precision simulations.

\section{Abelian dominance in local unitary gauges.}
In local gauges, we know quantum fluctuations are very large and 
there are huge noise. 
On the other hand, in non-local gauges like MA gauge, vacuum fluctuations are small and  we have been able to observe physical quantities very clearly.
There are much difference of quantum fluctuations 
between MA gauge and local unitary gauges. Hence to extract physical signals in local unitary gauges,  high precision investigations are
necessary. But such a study has not been done so far.
We perform such high precision investigations in local unitary gauges 
using the multi-level method developed by L\H{u}scher and Weisz\cite{Luscher:2001up}.

\subsection{The multi-level method}

The multi-level method is a powerful noise reduction method\cite{Luscher:2001up}.
This method is effective only with respect to local operators.
For example it is very useful to measure Polyakov loop correlation functions.
The procedure is the following:
1. Lattice is divided into $N_s$ sublattices.
2. A sublattice average of operators is taken.
3. The link variables are replaced 
using gauge update except for the spatial links 
at the boundary of the sublattice.
(it is called internal update.)
4. Step 2, 3 are repeated until stable values 
for the sublattice average of operators are obtained.
Now we adopt this method in the Abelian projection using local unitary gauges.
Then gauge-fixing and the Abelian projection steps 
are implemented in each internal update.
And operators are constructed by Abelian link fields. As a gauge-fixing condition, we consider  local unitary gauges consistent with the  multi-level method. 

\subsection{Local unitary gauge.}

\begin{figure}[h]
\begin{center}
\includegraphics[height=3.5cm]{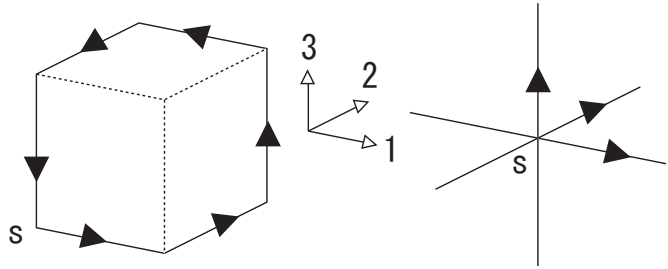}% Here is how to import EPS art
\caption{The left figure shows the cube operator diagonalized in  the F123 gauge.
The right  is the set of spatial Polyakov lines used in the SPL gauge.}\label{localug}
\end{center}
\end{figure}

We consider three local unitary gauges here. The first one is a
F12 gauge which makes 1-2 plane plaquette diagonal.
\begin{eqnarray}
G_{F12,n}\left(
U_{n,1}U_{n+\hat{1},2}U_{n+\hat{2},1}^{\dagger}U_{n,2}^{\dagger}
\right)G_{F12,n}^{\dagger}=\mbox{(diagonal)}.
\end{eqnarray}
The second is a F123 gauge  similar to the F12 gauge.
Its gauge-fixing matrix makes a 1-2-3 cube-like operator in Fig.\ref{localug} diagonal.
\begin{eqnarray}
G_{F123,n}\left(
U_{n,1}U_{n+\hat{1},2}U_{n+\hat{1}+\hat{2},3}
U_{n+\hat{2}+\hat{3},1}^{\dagger}U_{n+\hat{3},2}^{\dagger}U_{n,3}^{\dagger}
\right)G_{F123,n}^{\dagger}=\mbox{(diagonal)}.
\end{eqnarray}
The third is a spatial Polyakov line (SPL) gauge  similar to Polyakov gauge.
The gauge-fixing matrix makes spatial Polyakov line operators in Fig.\ref{localug} diagonal.
\begin{eqnarray}
G_{SPL,n}\left(
	 \sum_{\mu=1}^{3}\prod_{i=1}^{N_s}U_{n+(i-1)\hat{\mu},\mu}
\right)G_{SPL,n}^{\dagger}=\mbox{(diagonal)}.
\end{eqnarray}

\subsection{Numerical results}

We use a simple SU(2) Wilson gauge action.
We adopt the coupling constant $\beta=2.5$
which corresponds to the lattice spacing $a(\beta=2.5) = 0.082(2)[fm]$.
The lattice size is $24^4$.
For the multi-level method, the 
number of sublattices is 6, the
sublattice size is 4 and iterations of internal updates are over 80000.

\begin{figure}[h]
\begin{center}
\includegraphics[height=5.5cm]{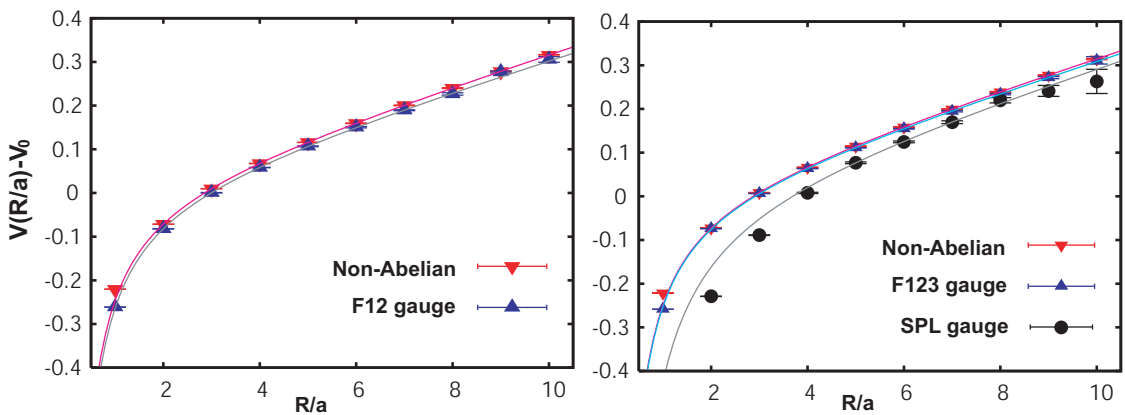}% Here is how to import EPS art
\caption {Left is a static 
potential in F12 gauge in comparison with the non-Abelian potential.
Right  is a potential in 
F123 gauge and SPL gauge in comparison with the non-Abelian potential.
The solid lines show a fitted function ($\sigma R -{c/R} + V_0$).}\label{multipote}
\end{center}
\end{figure}

First we show an Abelian static potential in F12 gauge
and compare it with non-Abelian one.
In the figure \ref{multipote}(left)  Abelian dominance is seen clearly.
This is a very interesting result.
Note that the Coulomb part is different from that in the non-Abelian potential. The Coulomb part is also different in MA gauge.
Next we show  in \ref{multipote}(right) Abelian potentials in  other two local unitary gauges.
The F123 gauge gives us almost similar results as in F12 gauge. On the other hand, the
SPL gauge gives  bigger  Coulomb coefficient. Nevertheless, 
explicit values of the fitted string tension  shown in Table \ref{stNAvsA} are the same within the statistical errors. 
We could conclude that Abelian dominance is confirmed also in these local unitary gauges. 

\begin{table}[h]
\caption[stringtension]{Fitted string tensions($\beta=2.5$).
The region for fitting is $R/a = 2 \sim 10$.}\label{stNAvsA}
\begin{center}
\begin{tabular}{c|cccc}
 & NonAbelian & F12 gauge & F123 gauge & SPL \\
\hline
String tension $\sigma$ & 0.034(2) & 0.034(2) & 0.034(2) & 0.034(2)\\
Coulomb coefficient $c$ & 0.29(1) & 0.29(1)  & 0.28(1)  & 0.65(3)
\end{tabular}
\end{center}
\end{table}

\section{Abelian dominance without gauge-fixing.}

\begin{figure}[h]
\begin{center}
\includegraphics[height=7.5cm]{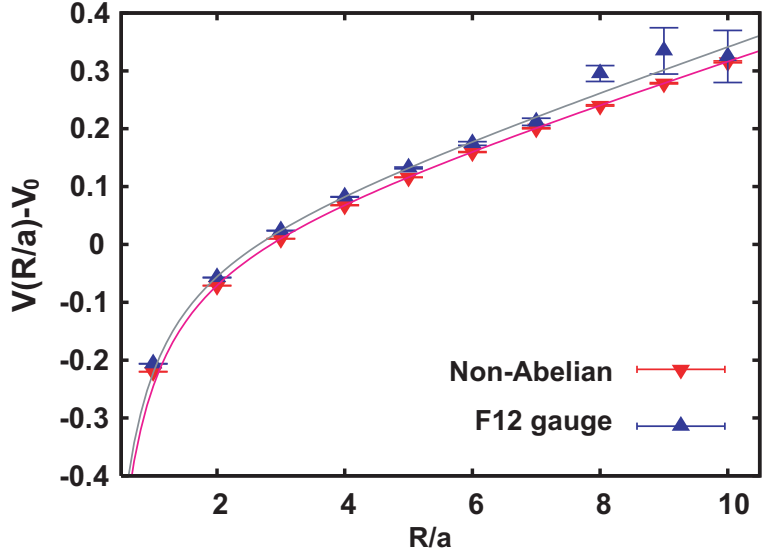}% Here is how to import EPS art
\caption{Abelian potential without gauge-fixing in comparison with the non-Abelian potential. The solid lines mean a fitted function ($\sigma R -{c/R} + V_0$).}\label{multipotengf}
\end{center}
\end{figure}

The Abelian dominance is found to be seen
in local unitary gauges where quantum fluctuations are large.
This suggests that Abelian dominance works in any Abelian projection scheme, although it is impossible to investigate  all gauge-fixing conditions. If it is true, gauge-fixing may not be essential
in Abelian confinement scenario.
To confirm this suggestion,
we measure an Abelian static potential and 
Abelian flux tube profiles without gauge-fixing.
In figure \ref{multipotengf}
we show the Abelian potential without gauge-fixing 
using the multi-level method.
The parameters for the multi-level method are the same 
as those used in local unitary gauge cases.
In no gauge-fixing case
Abelian and non-Abelian potentials are quite similar in all regions.
This results had been expected analytically by Ogilvie\cite{Ogilvie:1998wu}.
The fitted string tension $\sigma$ is 0.034(2).
This result shows Abelian dominance can be seen without any gauge-fixing.

\subsection{Abelian flux tube profiles}
\begin{figure}[h]
\begin{center}
\includegraphics[height=6cm]{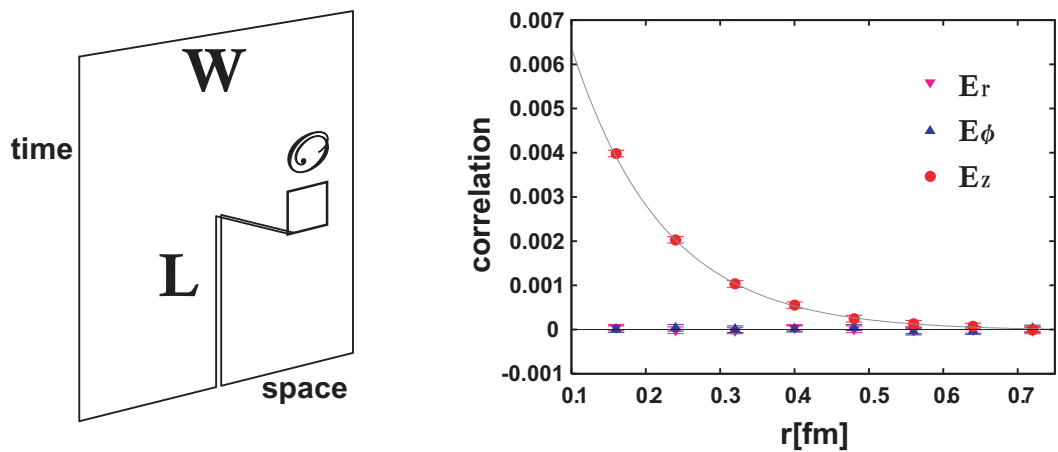}% Here is how to import EPS art
\caption{The left  shows a connected Wilson loop correlation function.
The right figure shows all components of the Abelian electric field.
These profiles are studied on a perpendicular plane 
at the midpoint between the two quarks. The
Wilson loops size is ($R\times T$)=($5\times 5$). }\label{connectP}
\end{center}
\end{figure}

Abelian flux tube profiles are important observables 
in understanding the Abelian scenario.
For this purpose we  calculate  correlations between the
Wilson loop and Abelian field strength without gauge-fixing.
However simple correlation functions between disconnected operators 
can not observe the profiles, since the Abelian plaquette 
operator is gauge-variant.
Here we measure  connected correlation functions 
to measure the Abelian flux tube profiles without gauge-fixing.
Similar investigations were done 
to measure non-Abelian electric fields around a pair of static quarks by Cea et al.
\cite{Cea:1995zt} and DiGiacomo et al.\cite{DiGiacomo:1989yp}.
The connected correlation function is defined as follows:
\begin{eqnarray}
\langle {\cal O}_A(r) \rangle_{W}&=&
{\langle {1\over 3}\sum_{a}\mbox{Tr}\left[
LW(r=0,R,T)L^{\dagger}\sigma^a{\cal O}_A^a(r)
\right]\rangle
\over \langle \mbox{Tr}\left[W(R,T)\right]\rangle}
\end{eqnarray}
where ${\cal O}_A$ is an Abelian field strength operator constructed by Abelian link fields, 
$L$ is a non-Abelian link field connecting the Wilson loop with the Abelian operator 
and $W$ is the Wilson loop (See figure \ref{connectP}).
To avoid the artificial effects of the bridge part $L$ and $L^{\dagger}$, we consider several types of L and then take an average of the results of all types adopted.

In this calculation 
we use the Iwasaki gauge action\cite{Iwasaki:1985we}
.
We adopt the coupling constant $\beta= 1.20$
which corresponds to the lattice distance $a(\beta=1.20)= 0.0792(2)[fm]$.
The lattice size is $32^4$.
After 5000 thermalizations, 
we have taken 4000 thermalized configurations
per 200 sweeps for measurements. 
To get a good signal-to-noise ratio, 
the APE smearing technique is used 
for evaluating Wilson loops\cite{Albanese:1987ds}.

We measured  all components of the Abelian electric fields $E_{A i}$.
\begin{eqnarray}
E_{A i} &=& \Theta_{n,4,i} \\
\Theta_{n,\mu,\nu} &=& \theta_{n,\mu}+\theta_{n+\hat{\mu},\nu}
-\theta_{n+\hat{\nu},\mu}-\theta_{n,\nu}
\end{eqnarray}
In figure \ref{connectP} 
we show that the Abelian electric fields are squeezed.
We try to fit using the  following  function:
\begin{eqnarray}
f(x)&=&c_1\exp(-r/\lambda)+c_0.
\end{eqnarray}
Here $\lambda$ is the penetration length
which is an important quantity in the Abelian dual Meissner effect.
The fitted penetration length $\lambda$ is 0.12(2)[fm].
This result is  consistent with that in  MA gauge 
in Ref \cite{Chernodub:2005gz}.
We find also that 
all components of the magnetic fields
have no correlation with the Wilson loop.
These results are consistent with the Abelian scenario.

\section{Conjecture and conclusions}

We have studied the gauge dependence problem with respect to 
Abelian confinement scenario. First we have studied extensively Abelian
projections in local unitary gauges with the help of the multi-level noise reduction method.
\begin{itemize}
\item In local unitary gauges (F12, F123, SPL)
Abelian dominance with respect to the string tension 
is observed very beautifully.
The difference is seen only in the Coulomb region.
\end{itemize}

Next we have investigated if the Abelian confinement scenario can be seen without any gauge-fixing. 
\begin{itemize}
\item 
Abelian dominance of the string tension 
is observed in all regions without any gauge-fixing.
\item 
Abelian electric fields are squeezed and the penetration length observed is consistent with that in MA gauge.
Here we have used many vacuum configurations using the Iwasaki improved gauge action on $32^4$ lattice.

\end{itemize}
These results suggest that
Abelian confinement scenario works without resort to Abelian
projection. 

On the basis of the above results, 
we propose a conjecture of a gauge-independent Abelian confinement
scenario;\\
{\bf Each Abelian component in any color direction 
becomes dominant in the infrared region.
Abelian dominance and the Abelian dual Meissner effect
are seen in any color direction. 
A state which is Abelian neutral in any color direction 
can be a physical state. It is just a color-singlet state.
In this way, color confinement can be understood 
in the framework of Abelian scenario in a gauge-invariant way.}\\
It is very interesting to confirm this conjecture numerically. The investigation is in progress.
%\newpage

\begin{flushleft}{\bf Acknowledgments}\end{flushleft}

The authors would like to thank Y. Koma 
for his simulation code of the multi-level method.
The numerical simulations of this work were done using 
NEC SX7 in RIKEN and NEC SX5 in RCNP. 
The authors would like to thank RIKEN and RCNP for their
support of computer facilities.

\end{document}